\documentclass[twocolumn,showpacs,showkeys,preprintnumbers,amsmath,amssymb]{revtex4}

\usepackage{graphicx}
\usepackage{dcolumn}
\usepackage{bm}

\begin{document}


\title{Phase Separation of Binary Systems}

\author{Tian Ma}
\affiliation{Department of Mathematics, Sichuan University,
Chengdu, P. R. China%
}%

\author{Shouhong Wang}
 \homepage{http://www.indiana.edu/~fluid}
\affiliation{Department of Mathematics,
Indiana University, Bloomington, IN 47405
}%
\thanks{This work is supported in part by grants from ONR and NSF.}
\newcommand{\R}{\mathbb R}
\newcommand{\cC}{\mathcal C}
\newcommand{\C}{\mathbb C}

\newtheorem{thm}{Theorem}
\newtheorem{cor}{Corollary}
\newtheorem{defi}{Definition}
\newtheorem{ex}{Example}
\newtheorem{lem}{Lemma}
\newtheorem{rem}{Remark}

\def\bt{\begin{thm}}
\def\et{\end{thm}}

\def\bl{\begin{lem}}
\def\el{\end{lem}}
\def\la{\label}

\def\bd{\begin{defi}}
\def\ed{\end{defi}}

\def\bc{\begin{cor}}
\def\ec{\end{cor}}

\def\bp{\begin{proof}}
\def\ep{\end{proof}}

\def\br{\begin{remark}}
\def\er{\end{remark}}

\date{\today}

\begin{abstract}
In this Letter, three physical predictions on the  phase separation of binary systems are derived based on  a dynamic transition theory developed recently by the authors. First, the order of  phase transitions is precisely determined by the sign of a parameter $K_d$ (or a nondimensional 
parameter $K$) such that if $K_d>0$, the transition is first-order with latent heat and if $K_d <0$, the transition is second-order. Second, a theoretical transition diagram is derived, leading in particular  to  a prediction that there is only second-order transition for molar fraction near $1/2$. This is different from the prediction made by the classical transition diagram. Third, a critical length scale $L_d^c$ is derived such that no phase separation occurs at any temperature if the length of the container is smaller than the critical length scale.
\end{abstract}

\pacs{05.70.Fh, 64.60.Ht}
\keywords{binary system, Cahn-Hilliard equation, phase diagram, order of separation, critical length scale, dynamic transition theory}
\maketitle

Materials compounded by two components $A$ and $B$, such as  binary
alloys, binary solutions and polymers, are called binary
systems. Sufficient cooling of a binary system may lead to phase
separations, i.e., at the critical temperature, the concentrations
of both components $A$ and $B$ with homogeneous distribution undergo  changes,
leading to heterogeneous spatial  distributions. 
The main objective of this Letter is to precisely describe the phase separation mechanism and to make a few physical predictions.

\medskip

{\bf Cahn-Hilliard Equation.} Let $u_A$   and $u_B$ be the concentrations of components $A$  and  $B$ respectively,  then $u_B=1-u_A$.  In a homogeneous state, $u_B=\bar{u}_B$ is a constant. We take $u$  to be  the concentration density deviation $u=u_B-\bar{u}_B.$
The Cahn-Hilliard free energy is given by
\begin{equation}
F(u)=F_0+\int_{\Omega} \Big[\frac{\mu}{2}|\nabla u|^2+ f(u)\Big]dx,\label{8.49}
\end{equation}
where  
$$f(u)= \alpha_1 u^2+\alpha_2 u^3+\alpha_3 u^4.$$
The same results in this article can be derived in the same fashion, and for simplicity, we take this form of $f$ as given here. Then the classical Cahn-Hilliard equation is as follows:
\begin{equation}
\left.
\begin{aligned} 
&\frac{\partial u}{\partial
t}=-k\Delta^2u+\Delta [  b_1 u^1+ b_2 u^2+ b_3 u^3     ],\\
&\int_{\Omega}u(x,t)dx=0,
\end{aligned}
\right.\label{8.57}
\end{equation}
supplemented with the Neumann boundary condition:
\begin{equation}
\frac{\partial u}{\partial n}=\frac{\partial\Delta u}{\partial
n}=0  \ \ \ \ \text{on}\
\partial\Omega ,\label{8.53}
\end{equation}
where   $\Omega=\Pi^3_{k=1}(0,L_k) \subset \R^3$   is  a rectangular  domain. 
We note that the more general domain case can be studied as well.

To derive  the nondimensional form of equation, let
\begin{equation}
\begin{aligned}
&x=lx^{\prime}, &&  t=\frac{l^4}{k}t^{\prime}, && u=u_0u^{\prime},\\
&\lambda =-\frac{l^2b_1}{k},&& \gamma_2=\frac{l^2b_2u_0}{k},
&& \gamma_3=\frac{l^2b_3u^2_0}{k},
\end{aligned}\la{nondim}
\end{equation}
where $l$ is a given length,  $u_0=\bar{u}_B$ is the constant
concentration of $B$, and $\gamma_3>0$. Then the equation (\ref{8.57}) can be
rewritten as follows (omitting the primes)
\begin{equation}
\left.
\begin{aligned} 
&\frac{\partial u}{\partial
t}=-\Delta^2u-\lambda\Delta u+\Delta (\gamma_2u^2+\gamma_3u^3),\\
&\int_{\Omega}u(x,t)dx=0,\\
&u(x,0)=\varphi .
\end{aligned}
\right.\label{8.58}
\end{equation}

{\bf Criteria of separation order.} Each 3D  rectangular domain is one of the following two cases:
 \begin{align*}
 & \text{ Case I:  } \quad L=L_1>L_j\qquad  \forall  j\leq 2, 3, \\
 & \text{ Case II: }\quad  L=L_1=L_2 > L_3 \text{ or }  L_1 = L_2 = L_3.
 \end{align*}
 We define a nondimensional parameter: 
\begin{equation}
K=\left\{
\begin{aligned}
& \frac{2L^2}{9\pi^2}\gamma^2_2 -  \gamma_3 
    &&  \text{ for Case I},\\
& \frac{26L^2}{27\pi^2}\gamma^2_2 - \gamma_3 
   &&  \text{ for Case II}.
   \end{aligned}
   \right.
\end{equation}
which, by (\ref{nondim}),  is equivalent to the following dimensional parameter:
\begin{equation}
K_d=\left\{
\begin{aligned}
& \frac{2L^2_d}{9\pi^2}\frac{b_2^2}{k}-  b_3  && \text{ for Case I}, \\
& \frac{26L_d^2}{27\pi^2}\frac{b_2^2}{k}  - b_3  && \text{ for Case II}.
   \end{aligned}
   \right.
\end{equation}
where $L_d=L \cdot l$  is the dimensional length scale.

By theorems proved in \cite{MW08d}, the order of transitions is determined by the sign of this parameter $K$ or $K_d$ as follows, and we have readily derived the following physical predictions:

\medskip

\noindent
{\sc Physical Conclusion I:} {\it The order of phase separation is completely determined by the sign of the nondimensional parameter $K_d$ as follows:

\begin{itemize}
\item[(1)] If $K_d<0$, the separation is second order and the dynamic behavior of the Cahn-Hilliard system is as shown in Figure~\ref{f8.15-1}.

\item[(2)] If $K_d>0$, the separation is first order transition with latent heat. In particular, there are two critical temperature $T^*> T_c$ such that if  the temperature $T> T^*$, the system  is in the homogeneous state, when $T^*> T > T_c$, the system is in metastable state accompanied with hysteresis corresponding to saddle-node  bifurcation, and when  $T< T_c$, the system is under phase separation state. In addition, the critical temperatures are functions of $u_0$  and $L$: $T^*=T^*(u_0, L),  T_c=T_c(u_0, L)$. See
 Figure~\ref{f8.16-1}.
\end{itemize}
}
\begin{figure}[hbt]
  \centering
  \includegraphics[width=0.4\textwidth]{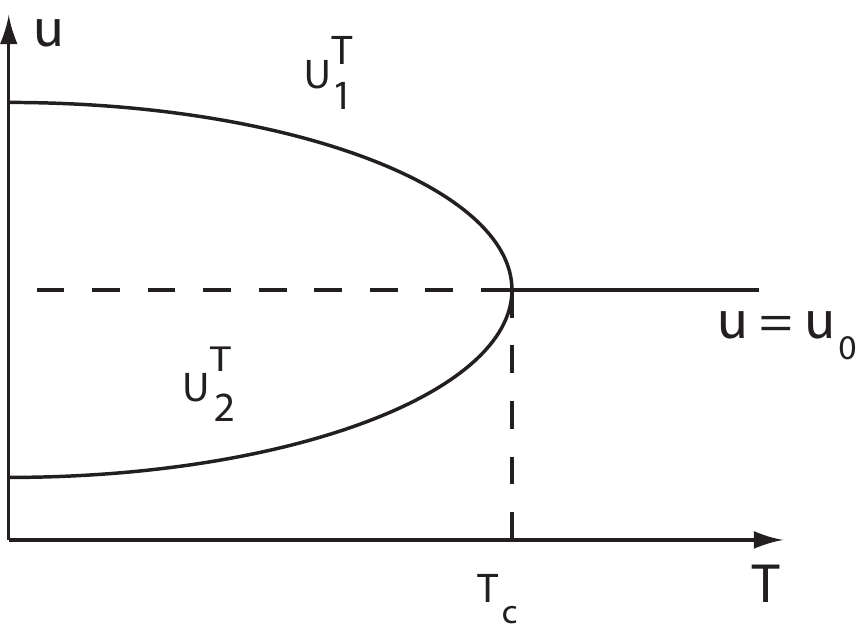}
  \caption{The state $u_0=\bar{u}_B$ is
stable  if $T_c < T$, and  the state $u_0$ is unstable, $U_1^T$ and $U^T_2$ are stable if $T < T_c$.}
  \la{f8.15-1}
 \end{figure}

\begin{figure}[hbt]
  \centering
  \includegraphics[width=0.4\textwidth]{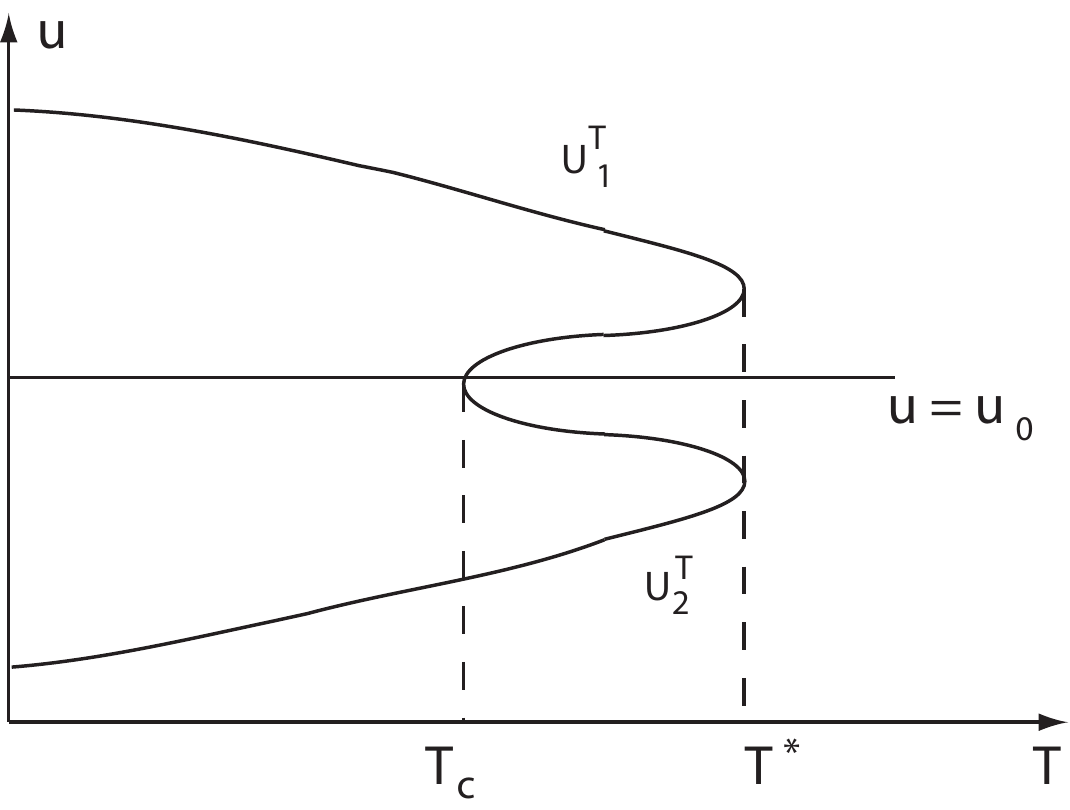}
  \caption{For fixed $u_0$  and $L$, the transition for the case where $K_d >0$ is  first order separation  with latent heat and with hysteresis: $U_1^T$  and $U_2^T$   represent separation states, and $u_0$ is the homogeneous state. In this case, for $T_c < T < T^*$, all states $u_0$, $U^T_1$, $u_2^T$  are metastable states. For $T< T_c$, $u_0$ is unstable, and $U^T_1$ and $U^T_2$ are stable states.}
\la{f8.16-1}
 \end{figure}
 This is in agreement with {\it part of} the classical phase diagram from the classical thermodynamic theory given in Figure~\ref{f8.14-2}; 
see, among others,  Reichl \cite{reichl}, 
 Novick-Cohen and Segal \cite{NS84} ,  and Langer \cite{langer71}. However, as we shall see below, our result shows that near $u_0=1/2$, there is no metastable region; see Figure~\ref{fch-4}.

 \begin{figure}[hbt]
  \centering
  \includegraphics[width=0.4\textwidth]{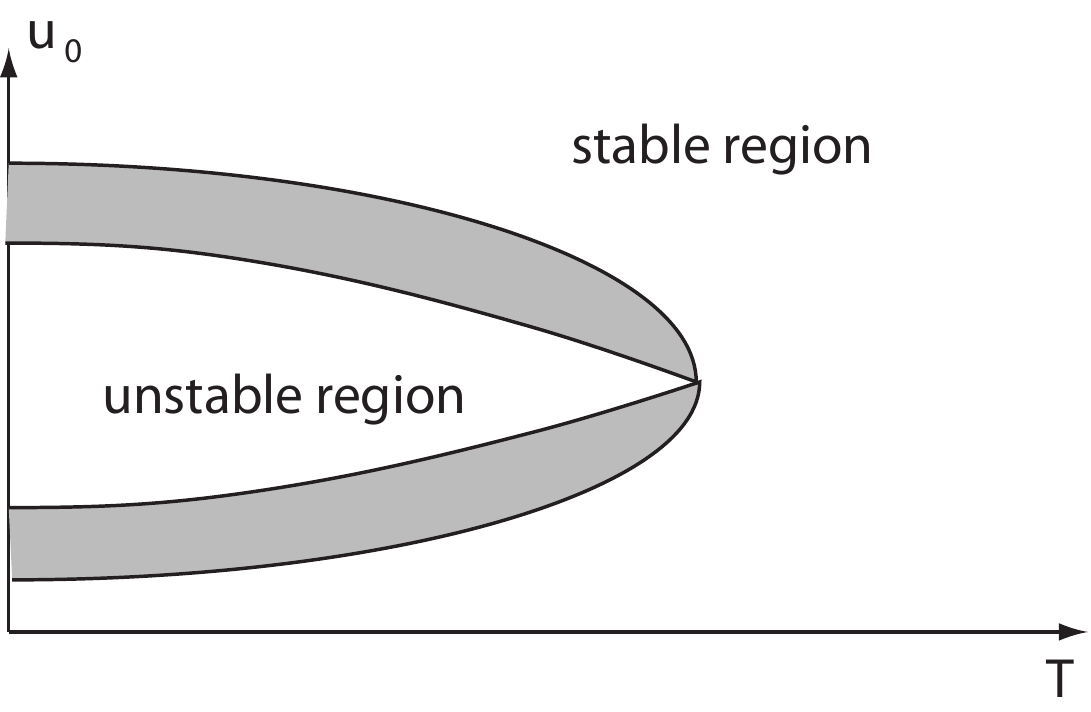}
  \caption{Typical phase diagram from classical
thermodynamic theory with shadowed being the metastable region .}
  \la{f8.14-2}
 \end{figure}

 \bigskip
 
 {\bf Transition diagram.}
We now examine the order of separation in terms of the length scale $L_d$ and mol fraction  $u_0$.  For this purpose, according to the Hildebrand theory (see Reichl \cite{reichl}), $b_2$  and $b_3$ can be expressed in two explicit formulas. Disregarding  the term $|\nabla u|^2$, the molar Gibbs free energy
takes the following form
\begin{align}
f=& \mu_A(1-u)+\mu_Bu+RT(1-u)\ln (1-u)\nonumber \\
& +RTu\ln
u+au(1-u),\label{8.122}
\end{align}
where $\mu_A,\mu_B$ are the chemical potential of $A$ and $B$
respectively, $R$ the molar gas constant, $a>0$ the measure of
repel action between $A$ and $B$. Therefore, the coefficients $b_2$  and $b_3$  are given by 
\begin{equation}
\label{ch-1}
\begin{aligned}
& b_2=\frac{D}{3 !} \frac{d^3 f(u_0)}{du^3}= \frac{2u_0-1}{6u^2_0(1-u_0)^2}D RT,\\
& b_3=\frac{D}{4 !} \frac{d^4 f(u_0)}{du^4}= \frac{1-3u_0 +3u^2_0}{12u^3_0(1-u_0)^3}DRT, 
\end{aligned}
\end{equation}
where $D$ is the diffusion coefficient.
It is easy to see that
\begin{align*}
&b_2\left\{
  \begin{aligned} 
     & =0&& \text{ if } u_0=\frac{1}{2},\\
&   \neq 0&& \text{ if } u_0\neq\frac{1}{2},
\end{aligned}
\right.\\
&b_3> 0 \qquad   \forall 0<u_0<1.
\end{align*}
It is clear that the above formulas for $b_2$  and $b_3$ based on the Hildebrand theory fail near  $u_0=0, 1$. However, the physically relevant case is away from these two end points of $u_0$, and then we have:
\begin{equation}
\begin{aligned}
& b_2 =\frac{16 DRT}{3} (u_0-\frac{1}{2}) + o(u_0-\frac12), \\
& b_3 =\frac{4DRT}{3} +  o(1).
\end{aligned}\label{ch-2}
\end{equation}
Then solving $K_d=0$ gives a critical (dimensional)  length scale $L_d$:
\begin{equation}
L_d = \left\{
\begin{aligned}
&\frac{3 \pi \sqrt{k}}{\sqrt{2}} \frac{\sqrt{b_3}}{|b_2|} && \text{ for Case I},\\
&\frac{3 \pi \sqrt{3k}}{\sqrt{26}} \frac{\sqrt{b_3}}{|b_2|} && \text{ for Case II}.
\end{aligned}\right. \label{ch-3}
\end{equation}
By (\ref{ch-2}) and (\ref{ch-3}), we have 
\begin{equation}
L_d = \left\{
\begin{aligned}
&\frac{3\sqrt{3k}\pi }{8\sqrt{2DRT_c}|u_0-\frac12|}   + O(1) && \text{for Case I},\\
&\frac{9 \sqrt{k}\pi}{8\sqrt{26DRT_c}|u_0-\frac12|}  + O(1) && \text{for Case II},
\end{aligned}\right. \label{ch-4}
\end{equation}
where $T_c$ is the critical temperature as given in Physical Conclusion I. 
From this formula, we derive the transition diagram given by Figure~\ref{fch-2}, and consequently, we derive a theoretical phase diagram given in Figure~\ref{fch-3}. 
In particular, we have shown the following physical conclusions:

\medskip

\noindent
{\sc Physical Conclusion II.} 
{\it 
\begin{itemize}

\item[(1)] For a fixed length scale $L=L'$, there are numbers $x_1 <\frac12< x_2$ such that 
the transition is second-order if the molar fraction $x_1< u_0< x_2$, and the 
transition is first-order   if $u_0 > x_2$ or $u_0 < x_1$. 

\item [(2)] The phase diagram Figure~\ref{fch-3} is for this  fixed length scale  $L'$. The points $x_1$  and $x_2$ are the two molar concentrations where there is no metastable region and no hysteresis phenomena for $x_1<u_0< x_2$. In other words, 
$$T^*(u_0)=T_c(u_0)  \qquad \text{ for } x_1 <u_0< x_2.$$
\end{itemize}
}

\medskip

 \begin{figure}[hbt]
  \centering
  \includegraphics[width=0.4\textwidth]{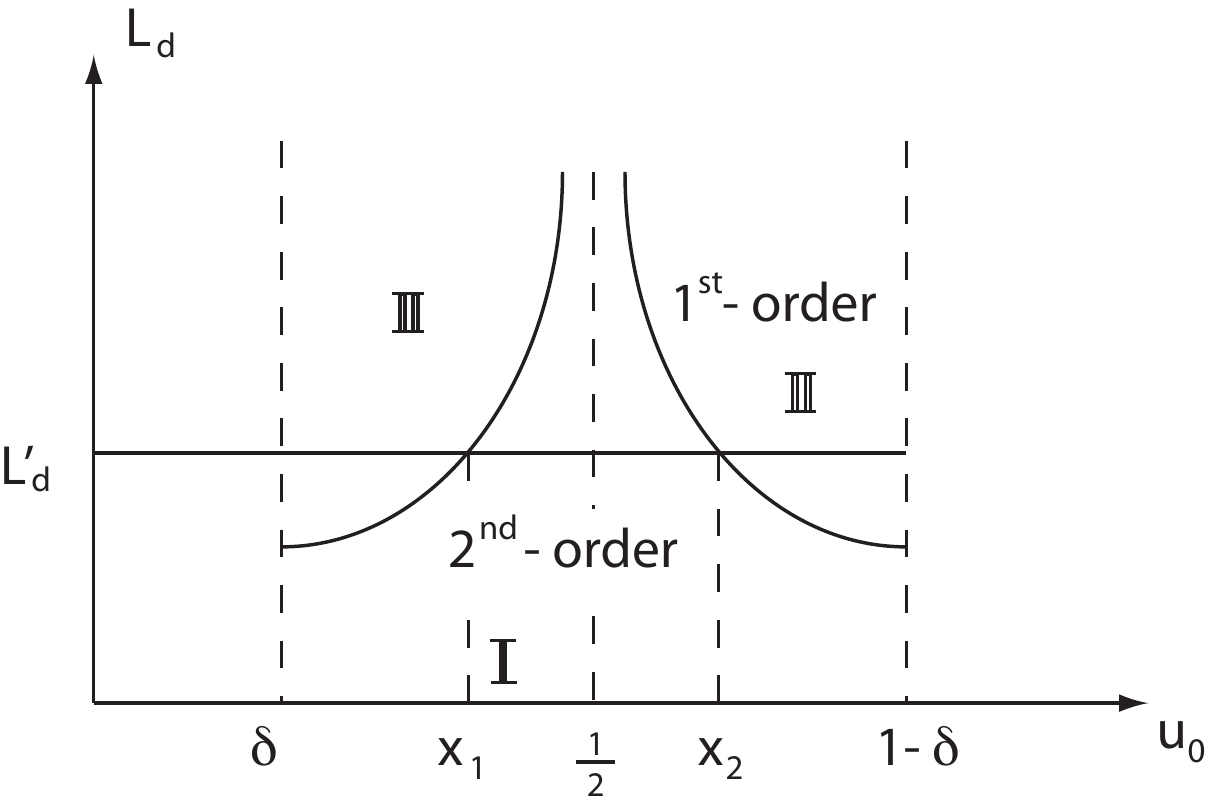}
  \caption{Transition diagram: region II is the first order transition region with latent heat, and region I is the second order transition region.}
  \la{fch-2}
 \end{figure}

 \begin{figure}[hbt]
  \centering
  \includegraphics[width=0.4\textwidth]{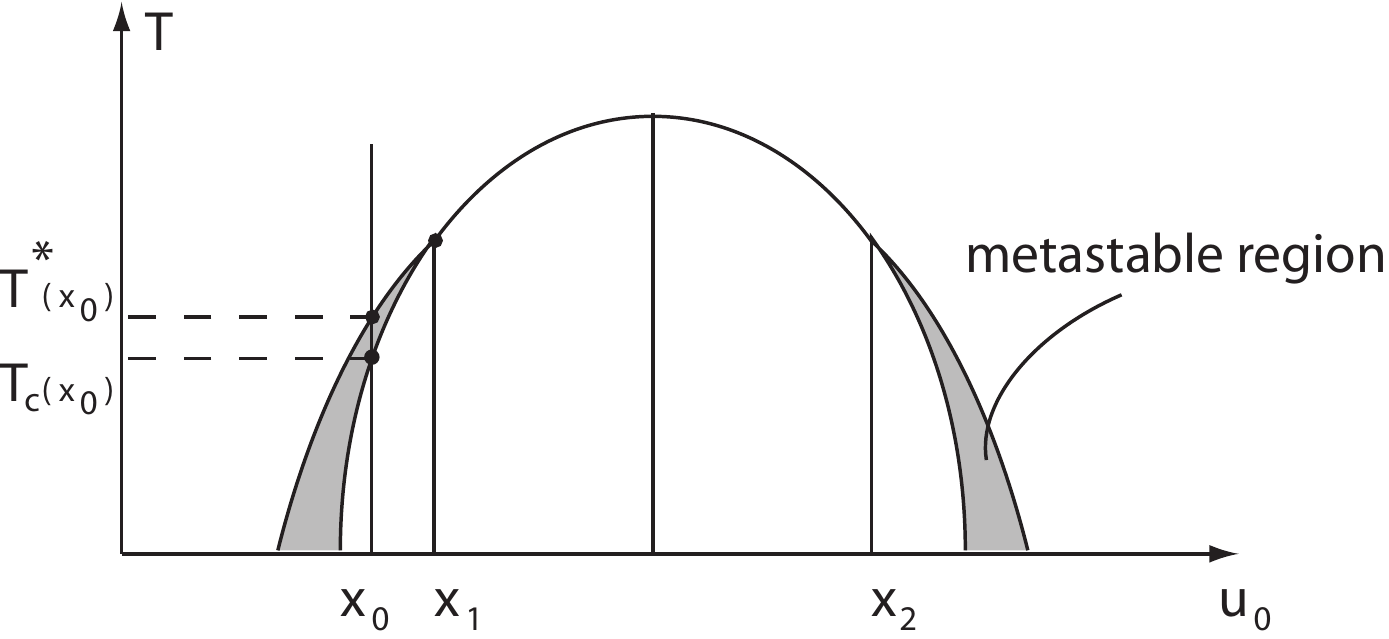}
  \caption{Derived theoretical diagram for a fixed length scale $L'$ with the shadowed region being the metastable region.}
  \la{fch-3}
 \end{figure}

 {\bf $TL$-phase diagram.} We now derive the length and temperature phase diagram. 
 For this purpose, we consider the linear eigenvalue problem for the Cahn-Hilliard equation as follows:
 \begin{equation}
\begin{aligned}
& - \Delta^2 u - \lambda \Delta u = \beta u, \\
& \frac{\partial u}{\partial n}=\frac{\partial\Delta u}{\partial
n}=0  \ \ \ \ \text{on}\
\partial\Omega.
\end{aligned}
\end{equation}
The first eigenvalue is given by 
$$ \beta_1 = - \frac{\pi^2}{L^2}\left( \frac{\pi^2}{L^2} - \lambda\right)   = 
- \frac{\pi^2}{L^2}\left( \frac{\pi^2}{L^2}  + \frac{\l^2 b_1}{k}\right).
$$
By (10), we have 
$$ b_1 = \frac{D}{2} \frac{d^2 f(u_0)}{du^2} = \frac{DRT}{2 u_0(1-u_0)} -\frac{a}{2}.
$$
The critical parameter curve equation $\beta_1=0$ is given by 
\begin{align}
T_c= & \frac{u_0(1-u_0)}{RD}\left( a- \frac{k\pi^2}{2l^2 L^2}\right)  \nonumber \\
=
& \frac{u_0(1-u_0)}{RD}\left( a- \frac{k\pi^2}{2L_d^2}\right).\label{critical-t}
\end{align}
Using this formula  and the theorems in \cite{MW08d}, we derive the $TL$ phase 
diagram given by Figure~\ref{fch-4}, and the following physical conclusions:
\begin{figure}[hbt]
  \centering
  \includegraphics[width=0.4\textwidth]{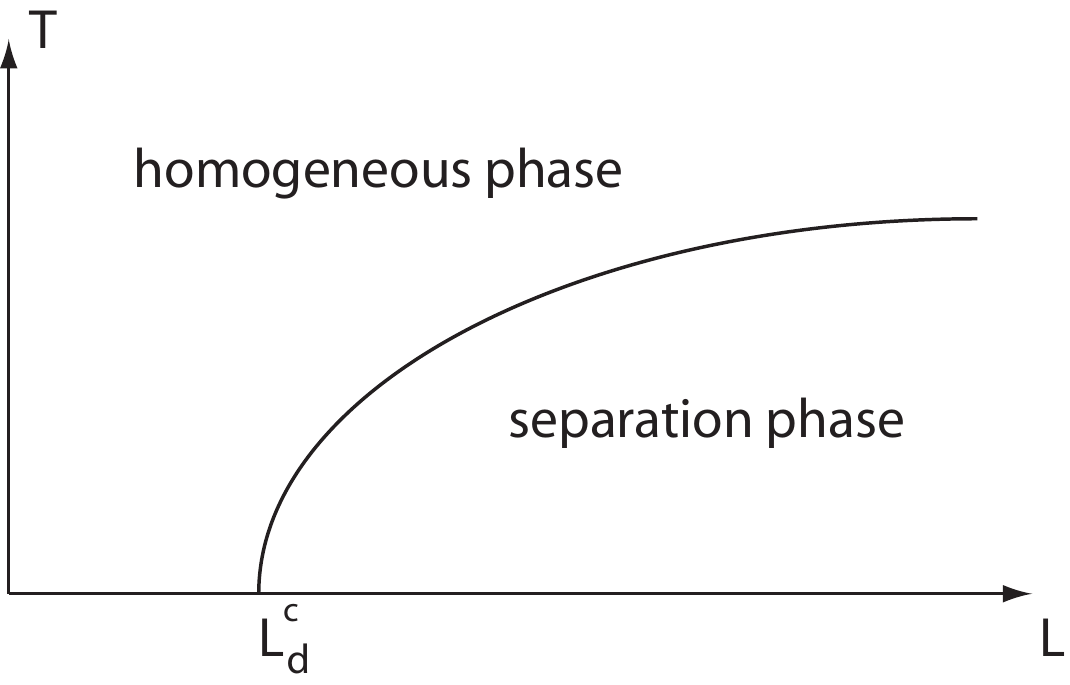}
  \caption{TL phase diagram.}
  \la{fch-4}
 \end{figure}
 
 \medskip

\noindent
{\sc Physical Conclusion III.} 
{\it 
For a given molar fraction $0<u_0 < 1$, there is a critical (dimensional) length 
$$L^c_d= \sqrt{\frac{k\pi^2}{2 a}}$$
such that the following hold true:
\begin{itemize}

\item[(1)] 
For $L_d < L^c_d$, there is no phase separation  for any temperature.

\item[(2)] For  $L_d  > L^c_d$, phase separation occurs at the critical temperature 
$T=T_c$  given by (\ref{critical-t}).
\end{itemize}
}

\medskip

{\bf Summary.} Based on a  dynamic transition theory developed recently by the authors 
\cite{b-book, chinese-book,MW08c, MW08f},  a systematic mathematical analysis is made for the Cahn-Hilliard equation modeling phase separation of binary systems \cite{MW08d}. 
Based on this rigorous analysis, we are able to make  three physical predictions on the  phase separation of binary systems: 

{\sc First}, the order of  phase transitions is precisely determined by the sign of a parameter $K_d$ (or a nondimensional  parameter $K$) such that if $K_d>0$, the transition is first-order with latent heat and if $K_d <0$, the transition is second-order. This parameter $K_d$  is explicitly given in terms of the 
system properties and the geometry of the container.

{\sc Second}, a theoretical transition diagram is derived, leading in particular  to  a prediction that there is only second-order transition for molar fraction near $1/2$. This is different from the prediction made by the classical transition diagram. 

{\sc Third}, a critical length scale $L_d^c$ is derived such that no phase separation even occurs at any temperature if the length scale of the container is smaller than the critical length scale. The transition temperature $T_c$ is precisely given  as well for the length scale is larger than the critical scale.

{\sc Finally}, our theory fully reveals the transition dynamics.  This is the advantage of using the dynamic classification scheme as proposed in \cite{chinese-book,MW08c, MW08f}, where the transitions are classified as Type-I, Type-II and Type-III. Also, we would like to mention that our results are derived for rectangular domains, and more general domain case can be studied using the dynamic transition theory as well, and other transition types such as the mixed transition may occur; see \cite{MW08d}.

\bibliographystyle{siam}

\def\cprime{$'$}

\end{document}